\documentclass[aps,prb,preprint,superscriptaddress]{revtex4-1}
\newcommand{\beaa}{\begin{eqnarray*}}
\newcommand{\eeaa}{\end{eqnarray*}}
\newcommand{\bea}{\begin{eqnarray}}
\newcommand{\eea}{\end{eqnarray}}
\newcommand{\be}{\begin{equation}}
\newcommand{\ee}{\end{equation}}
\def\ra{\rightarrow}
\usepackage{graphicx}
\usepackage{dcolumn}
\usepackage{bm}
\usepackage{amssymb}

\begin{document}

\markboth{Saarela,Kusmartsev}
{Doping induced electronic phase separation and Coulomb bubbles in layered superconductors}

%
%

\title
{DOPING INDUCED ELECTRONIC PHASE SEPARATION AND COULOMB BUBBLES 
IN LAYERED SUPERCONDUCTORS}

\author{M. Saarela$^*$}

\address{Department of Physical Sciences ,University of Oulu,\\
P.O.Box 3000, FIN-90014 University of Oulu, Finland\\
$^*$E-mail: Mikko.Saarela@oulu.fi}

\author{F V Kusmartsev}
\address{Department of
Physics, Loughborough University, Loughborough LE11 3TU, UK\\
E-mail: F.Kusmartsev@lboro.ac.uk}

\begin{abstract} We study properties of charge fluids with random impurities or heavy polarons using a microscopic Hamiltonian with the full many-body Coulomb interaction. At zero temperature and high enough density the bosonic fluid is superconducting, but when density decreases the Coulomb interaction will be strongly over-screened and impurities or polarons begin to trap charge carriers forming bound quasiparticle like clusters, which we call Coulomb bubbles or clumps. These bubbles are embedded inside the superconductor and form nuclei of a new insulating state. The growth of a bubble is terminated by the Coulomb force.  The fluid contains two groups of charge carriers associated with free and localized states.  The insulating state arises via a percolation of the insulating islands of bubbles, which cluster and prevent the flow of the electrical supercurrent through the system.  

Our results are applicable to HTSC. There the Coulomb fluids discussed in the paper correspond to mobile holes located on Cu sites and  heavy polarons or charged impurities located on Oxygen sites in interlayers between  CuO planes. As a result of our calculations the following two-component picture of two competing orders ( superconducting and local CDW) in cuprates arise. The mobile and localized  states are competing with each other  and their balance is controlled by doping. At high doping a large Fermi surface is open. There the density of real charge carriers is significantly larger than the density of the doped ones. When doping decreases more and more carriers
are localized as Coulomb clumps which are creating around heavy polarons localized on Oxygen sites and forming a regular lattice. The picture is consistent with the Gorkov and Teitelbaum (GT) analysis \cite{34,GT} of the transport, Hall effect data and the ARPES spectra as well as with nanoscale superstructures observed in Scanning Tunneling Microscope(STM) experiments [3-8]. The scenario of the clump formation may be also applicable to pnictides, where two types of clumps may arise even at very high temperatures. The paper has been published in the book "Condensed Matter Theories" \cite{Kus-CMT32}
\end{abstract}


\maketitle
\pagenumbering{arabic}
\pagestyle{plain}

\section{Introduction}\label{intro}

Superconductivity is one of the most fundamental quantum phenomena in condensed matter physics. After discovery of high transition temperature in cuprates and more recently, for instance, in pnictides the field has become very exciting, because new electronic devices have become feasible, but also because these materials show unconventional behavior as superconductors. In the conventional BCS theory of superconductivity [\cite{14}] electrons are paired in momentum space, forming so-called Cooper pairs. Cooper pairs are bosons and can occupy a coherent, macroscopic Bose-condensate state. Thus at suffieciently low temperatures the system becomes superfluid and the superconductivity may then be described as the superfluid flow of the charged condensed liquid. Yet, the conventional Cooper pairing corresponds just to momentum space correlations between the motion of two electrons and in this sense they are not point-like bosons at all. In the system of charged boson particles the superconducting state arises in a similar way when a macroscopic number of bosons is condensed on the lowest possible energy level and becomes superfluid below a critical temperature. According to Landau theory [\cite{13}] the flow will be superfluid when its velocity is lower than the critical velocity associated with low energy elementary excitations. 

In unconventional superconductors like cuprates the size of Cooper pairs associated with the coherence length or with the value of the superconducting gap is much smaller than in the conventional ones. Precisely speaking it is inversely proportional to square root of the superconducting gap. 
The matter is that  the coherence length  and therewith the size of Cooper pairs in the optimally-doped cuprates is of the order of inter-particle distances. This is in contrast to low temperature superconductors where the size of Cooper pairs is usually of hundreds or even thousands of interelectronic distances.  This reasoning is probable breaking down in over-doped cuprates where the value of superconducting gap decreases  and therefore the "size" of Cooper pairs increases and becomes of the order of dozens of interparticle spacings. In the under-doped case the picture is much more complicated because of the pseudogap and the superconducting phase is very inhomogeneous. 

Increasing temperature  or disorder destroys the superconducting state. In this paper we show that the superconducting state vanishes even at zero temperature when the boson density decreases. The effect arises due to over-screening of the Coulomb interaction and  the presence of charged impurities. When the density decreases correlations between charge carriers become stronger and that induces over-screening of the Coulomb interaction between bosons and each charged  impurity. As a result a spontaneous formation of a cluster of self-trapped charge carriers around each impurity takes place, which we will call a {\it Coulomb bubble or clump} (CB). 

The mechanism of formation of Coulomb bubbles and electronic phase separation described here for the quantum charged bosons is equally applicable to fermionic charged particles, because in both cases the dominating role is played by the Coulomb interaction and its nonlinear screening, which arises when the charged density decreases.  Although, it is also important to note that for fermionic Coulomb bubbles the internal, energetical and spatial structure will be slightly different as well as the critical density when the CBs first appear. For fermions bound inside the Coulomb bubble the Pauli principle must be obeyed and lower density is required for the bubble formation. Yet, in that strongly correlated regime the kinetic electron energy (or the Fermi energy, $E_F$) is significantly smaller than the characteristic energy of the Coulomb interaction $E_c$. This ratio determines the parameter $r_s=E_c/E_F$. Here we concentrate to the physics in the under-doped region of cuprates where $r_s>>1$. There are many experimental indications that pairs of electrons (or, more precisely, holes) are bound in the coordinate space, although the nature of the pairing in HTSC at any doping is not yet established and many debates are still going on. 

Taking into account the ratio between the size of the Cooper pair and the inter-electron distance for the optimally doped case one may notice a very appealing possibility that in the under-doped region when doping decreases the size of pairs could decrease and therefore the type of pairing has crossed from BCS type to  BEC regime. In this case the applicability of the charged Bose fluid is valid for the homogeneous fluid unless the temperature is so high that pairs break and then we need to take into account the Fermi kinetic energy, which in a correlated regime is a small correction to the energy of the inter-particle Coulomb interaction. In the under-doped region the coherence length is small, ie of the order of Bohr radius. Even in this regime the pairs may break when bosons form clusters. Then the bosonic bubbles became fermionic bubbles while the physical picture of the Coulomb bubbles formation remains the same
because of the universality of the Coulomb interaction dictating the physics in the strong correlated regime. 

Therewith arguments  in mind we present here a fully microscopic many-body description and numerical simulations of the quantum charged  boson fluid that includes arbitrary charge fluctuations and disorder in the form of charged point impurities. The Coulomb fluids studied correspond to mobile holes located on Cu sites and  heavy polarons or charged impurities located on Oxygen sites. We start with original two-band model from which t-J model has been derived\cite{ZR-singlet}. As a result of our calculations the following two-componet picture of two competing orders in cuprates arise. The mobile and localized  states are competing with each other  and
there arises a transformation of one type of state into another one with the doping. The states are co-existing and their balance is controlled by doping. At high doping a large Fermi surface is open. 
There the density of real charge carriers (holes) is significantly larger than the density of the doped ones. 
The polarons are very heavy, they are  localized on oxygen sites and therefore do not contribute into transport. When doping decreases more and more carriers
are localized in a form of Coulomb clumps which are creating around heavy polarons localized on Oxygen sites. Finally in the limit of low doping the clumps are forming a regular "clump" lattice. 
In this paper we demonstrate explicitly how the over-screening leads to the spontaneous formation of Coulomb clumps or CBs  in layered or quasi-two-dimensional superconductors, when the density of current carriers decreases. The charge neutrality of the whole system is maintained by a charged, structureless jellium background, which is mostly associated with polarons localized on oxygen sites in Cu-O planes of cuprates. 

Although superconductivity has been predicted to persist even in the presence of impurities\cite{8}, many experiments performed on thin films have shown a transition from a superconducting to an insulating state with increasing concentration of impurities, which induce disorder. Here we show that SIT may also arise with decreasing doping, i.e. with decreasing disorder, when the concentration of both impurities and bosons (electrons or holes) decreases. There the Coulomb interaction between particles and screening effects play the primary role. With decreasing electron density the system undergoes an electronic phase separation transition into two phases superconducting and insulating. In order to destroy the superconducting state the CBs should form an infinite cluster. In a closer look we arrive at the following scenario of the quantum SIT. In the attempt to screen the individual charges of impurities, the local superconducting density develops a giant charge density oscillation (GCDO) around each impurity or a polaron localized on O site\cite{Kus-Saarela-2009}. The GCDO amplitude increases when the density of the superfluid boson decreases. An electronic phase separation into localized and superconducting states takes place when the amplitude at the first minimum around the absolute maximum of the GCDO is comparable with the average boson density. As a result a droplet of charged bosons is self-trapped around the impurity and decoupled from the rest of the superconducting condensate. The number of particles inside each droplet is fixed and therefore the phase is fluctuating, while in the rest of the condensate the phase is fixed and charge is fluctuating, giving the possibility to form a superconducting current.

\begin{figure}[tb] 
\centerline{
\includegraphics[width=15cm,height=10cm]{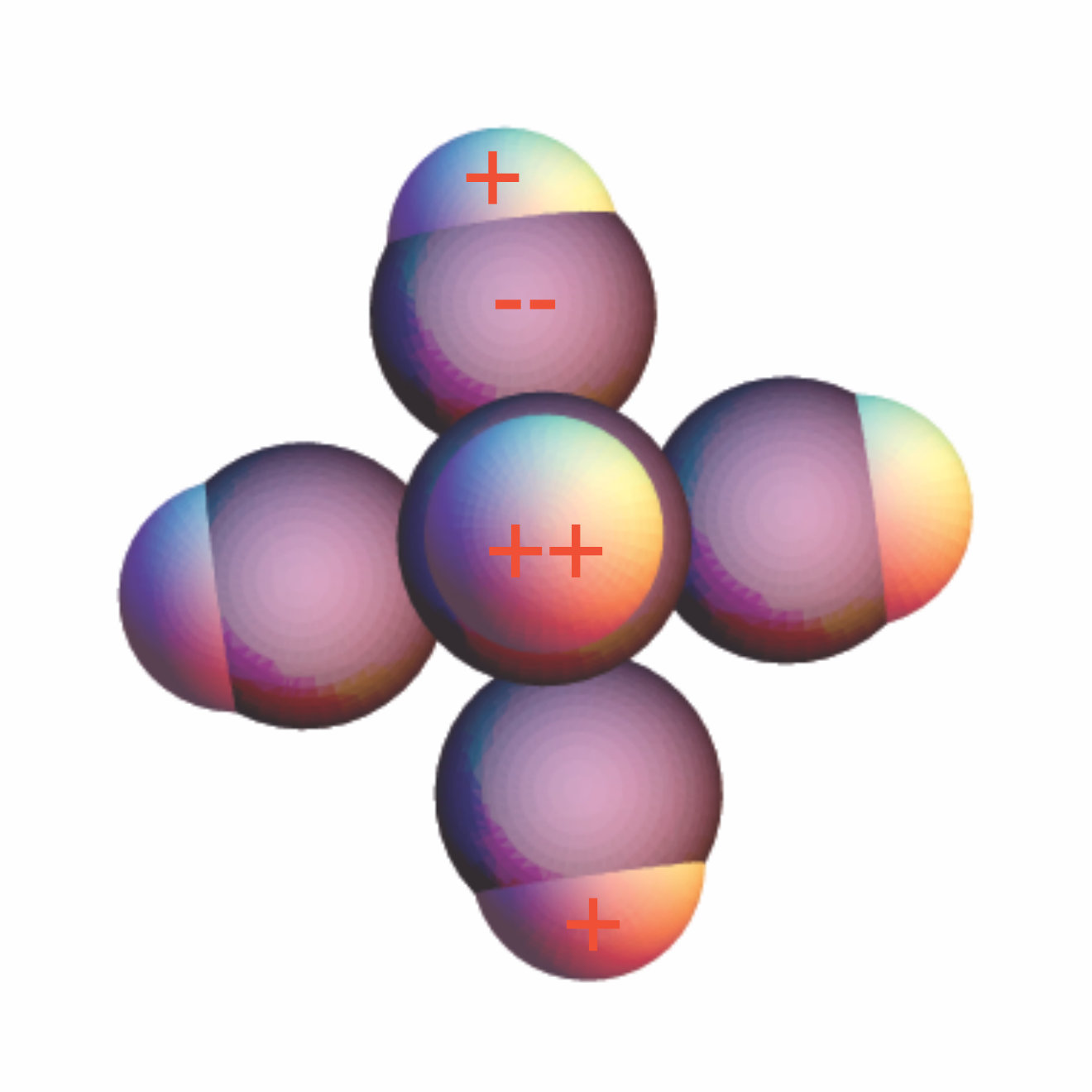}
}
\caption{(Color
online) A schematic structure of the Coulomb clump arising in charged quantum liquids.  A giant positive charge  arises just on the top or near of the negative impurity stimulated by over-screening. It may have a structure as two holes trapped with joint electrons around just on a single Cu-O plaquette. Such a cluster forms an elementary brick from which bigger clusters are formed.
} 
\label{Fig1} 
\end{figure}

The Coulomb clumps are microscopic molecules which are formed from heavy polarons accompanied  by  holes trapped due to many-body over-screening. For example on Fig.1 we present a clump consisting of a hole or two holes localized on Cu-site.
with 4 polarons trapped from neighboring Oxygen atoms. It may also have a more complicated structure, some kind of dump-bell shape as two holes attached to a polaron located on Oxygen site. Further polarons can attached to these two holes and so on. The localized electronic molecule behaves like a quantum dot under Coulomb blockade, but because of its small size the superconducting order parameter penetrates inside the bubble due to to the proximity effect \cite{9,10,11,12,Kus-Saarela-2009} . Such superconducting state with embedded heavy electronic electronic molecules or CBs behaves like a bulk superconductor and is a new kind of a ground state where the superfluid density is small and there are hidden nanoscale superstructures formed by these molecules. At very low temperatures these molecules are localized and therefore they can be revealed by STM.

\section{Variational theory of two-band model for HTSC compounds}

Original model for cuprate compounds, t-J model, has beed derived by Zhang and Rice in 1988 from a two-band model\cite{ZR-singlet}.  Since that time the majority of the theoretical research has been focused around this single band t-J model.
In their derivation  Zhang and Rice have started with the Hamiltonian describing a single
layer of square planar coordinated Cu and O atoms\cite{ZR-singlet} having the form:
\be
H=\sum_{i,\sigma} \epsilon_d d^{\dagger}_{i\sigma} d_{i\sigma}+\sum_{i,\sigma} \epsilon_p p^{\dagger}_{i\sigma} p_{i\sigma}+U \sum_{i} d^{\dagger}_{i\uparrow} d_{i\uparrow}  d^{\dagger}_{i\downarrow} d_{i\downarrow}+H'
\label{ZR-Hamiltonian}
\ee
where $H'$ is the hybridization between $d$ holes located on Cu sites and $p$ holes located on oxygen sites.
\be
H'=\sum_{i,\sigma} \sum_{l\in \{i\}} V_{il}d^{\dagger}_{i\sigma} p_{l\sigma} +H.C.
\label{hybrid}
\ee
where $V_{il}= \pm t$, ie  it is proportional to the overlapping integral $t$, see, the Ref.\cite{ZR-singlet}.
The sign depends on the sign of the overlapping integral.
When we neglect the on-site Coulomb interaction on Cu- sites, ie put $U=0$ we obtain the following energy spectrum consisting three branches:
\be
\epsilon_{\pm}(k_x,k_y)= \pm \sqrt{\epsilon_p^2/4+4t - 2 t\cos(k_x)-2 t\cos(k_y)},  {~\rm and~} \epsilon_0(k_x,k_y)= \epsilon_p/2
\label{spectrum}
\ee
where we put for simplicity $\epsilon_d=0$.
The spectrum indicates that there are two types of states here: localised (the flat band) and mobile, which may be associated with any of other two empty or partially filled bands.
ZR have considered a hole  localized  on a single Cu-site which is hybridized with hole located on four joint Oxygens. They have also shown that such state known now as ZR singlet may be mobile and
its behavior and many-body physics arising with the doping can be described with the aid of the t-J model\cite{ZR-singlet}. 

Our starting point is the same as had Zhang and Rice is that the Cu-O planes in cuprates are described by a similar two-band model associated with holes located on Cu and O sites. However to address the issue of localised states  observed recently in numerous experiments we would like to extend the ZR Hamiltonian to include the long-range Coulomb interaction. Indeed an inclusion of such Coulomb interaction made a dramatic difference namely in the appearance of the localised states associated with
an over-screening of the Coulomb repulsion\cite{EHLcluster,Kus-Saarela-2008,Kus-Saarela-2009} .
Here we would like to show that besides the mobile ZR singlets or free spinless holes on a square lattice, there may arise localised states, which energy levels are located below the hole empty band, which is filled with doping. These states arise due to
a nonlinear many-body screening. To show this and  to describe the coexistence of mobile and localized hole states we use an effective mass approximation. From the energy spectrum of the upper hole band, see, eq.(\ref{spectrum}), (we assume that this band will be doped) we derive an effective mass of holes equal to $m= \epsilon_p/4t$ and consider the quantum hole liquid in continuum approximation. Such approximation simplifies the theoretical treatment while keeping the essence of the Coulomb physics and Coulomb interaction. We also replace
the hole fermions by hole bosons since in the  small density limit the Coulomb interaction studied here is dominated.
The localized state may be induced by heavy impurities
or small heavy polarons which may be located on oxygen sites. Thus, one may argue  that the hole band is associated with with Cu- sites. While the second flat band may be associated to electronic small polarons localised on Oxygen sites.  The limit of the over-doped case is  most interesting. There established that  holes are forming a large Fermi surface, while very reasonable to have  electronic polarons localised on Oxygen sites. These polarons may act as heavy impurity as well and may trap holes due to
changing in the screening of the Coulomb interaction which arises with the doping. As a result
there will arise e-h strings or clumps which are embedded into a quantum hole liquid.
Thus, to describe the formation of clumps in these superconducting materials we start with the microscopic Hamiltonian where charged, random impurities or heavy polarons are embedded into the two-dimensional charged gas of current carriers,
\be
H_e + H_I=-\sum_{i=1}^{N}\frac{\hbar^2}{2m}\nabla_i^2 
+\frac 1 2 \mathop{{\sum}'}_{i,j=1}^{N}
\frac{e^2}{\varepsilon|{\bf r}_i-{\bf r}_j|}
-\frac{\hbar^2}{2M}\nabla_0^2 
-\mathop{{\sum}}_{i}^{N}
\frac{e^2}{\varepsilon|{\bf r}_0-{\bf r}_i|}\,.
\ee
We have separated the purely electronic (or bosonic) part $H_e$ from terms containing the impurity or a single heavy small polaron with the mass $M$,  and describing by the Hamiltonian, $H_I$. For simplicity we consider only one single impurity or a polaron. The number of mobile particles holes (or bosons) is $N$, they have the mass $m$, charge $e$ and position ${\bf r}_i$.  The impurity or heavy localised polaron is placed at ${\bf r}_0$ and its kinetic energy is controlled by the mass $M$. For a localized impurity or a polaron we let the mass grow to infinity (in the case of small polaron the effect is known as a collapse of the polaron band). We also add into the Hamiltonian a structureless, charged jellium, which neutralizes the system. The strength of the Coulomb interaction depends on the dielectric constant $\varepsilon$. Both the hole (or boson) mass and dielectric constant depend on the band structure of the material and it is convenient to use the atomic units where all distances are given in units of $r_0=r_s r_B$ with the Bohr radius $r_B$ and energies in Rydbergs. 

The  ground-state wave function which contains correlations between bosons and impurities is chosen in the form of   the Jastrow-type variational ansatz\cite{FeenbergBook}
\be
\Psi({\bf r}_1,{\bf r}_2,\ldots,{\bf r}_N)
	=e^{\frac1 2\sum^{N}_{i,j=1}
	u^{bb}(|{\bf r}_i,{\bf r}_j|)} 
	e^{\frac1 2\sum^{M}_{i=1}
	u^{It}(|{\bf r}_i,{\bf r}_0|)} 
	e^{\frac1 2\sum^{N}_{i=M+1}
	u^{In}(|{\bf r}_i,{\bf r}_0|)}
\ee
We have extended the conventional many-body variational theory to the case where $M$ bosons are trapped by the impurity. The wave function includes now a product of three components, boson-boson (bb) correlations, impurity-trapped boson (It) correlations and impurity non-trapped boson (In) correlations. Since particles are indistiquisable bosons the wave function should be symmetized with respect to trapped and non-trapped particles, but as we will show later the overlap of their distributions becomes vanishingly small and we can ignore the additional terms comming from the symmetrization.  

The key ingredients of the theory are, along with the correlation functions, the two-particle distribution functions,
\bea
 \rho^{bb}(|{\bf r}_1-{\bf r}_2|)
 &=& N(N-1)\frac{\int d^2r_3\ldots d^2r_N 
 \left|\Psi({\bf r}_1,\ldots,{\bf r}_N)\right|^2}
 {\langle\Psi| \Psi\rangle}
 \cr
 &=&n_0^2 g^{bb}(|{\bf r}_1-{\bf r}_2|)
\cr\cr
 \rho^{I}(|{\bf r}-{\bf r}'|)
  &=&\sum_{i=1}^N{\frac{\langle\Psi^{IM}\vert \delta({\bf r}_0-{\bf r})
 \delta({\bf r}_i-{\bf r}') \vert \Psi^{IM}\rangle}
		{\langle\Psi^{IM}\vert\Psi^{IM}\rangle}}
\cr
 &=& \frac{n_0}{\Omega}\left[  g^{It}(|{\bf r}-{\bf r}'|)
  +g^{In}(|{\bf r}-{\bf r}'|)\right]
\eea
where $g^{bb}(|{\bf r}_1-{\bf r}_2|)$, $g^{It}(|{\bf r}_0-{\bf r}_1|)$ and $g^{In}(|{\bf r}_0-{\bf r}_1|)$ are the boson-boson, impurity -trapped boson and impurity-non-trapped boson radial distribution functions, respectively, $\Omega$ is the total volume occupied by the system and $n_0$ is the boson density. 

The perfect screening condition requires the following normalizations, when we assume that the charge of the impurity, $-e$, has opposite sign to the boson charge. 
\bea
\int d{\bf r} \rho^{I}(r)=  \frac N{\Omega}
\cr
n_0\int d{\bf r} g^{It}(r)=M
\cr
n_0\int d{\bf r} [g^{In}(r)-1]= 1-M
\label{norm}
\eea
The Fourier transforms of the distribution functions define the static structure functions,
\bea
	S^{bb}(k)=1+n_0\int d^2r e^{i{\bf k}\cdot{\bf r}}[g^{bb}(r)-1]
	\cr
	S^{IM}(k)=n_0\int d^2r e^{i{\bf k}\cdot{\bf r}}[g^{It}(r)+g^{In}(r)-1]
	\,.
\eea
The additional information needed to solve variational many-body problem is the connection between the correlation functions and the physically observable distribution functions. This is provided by the infinite order hypernetted-chain (HNC) equations with the sum of nodal diagrams\cite{FeenbergBook}. 
\bea
g^{bb}(r)&=& e^{u^{bb}(r)+N^{bb}(r)}
\cr
g^{It}(r)&=& e^{u^{It}(r)+N^{I}(r)}
\cr
g^{In}(r)&=& e^{u^{In}(r)+N^{I}(r)}
\eea
Here we have made the essential assumption that in the impurity radial distribution functions the dependences on $\exp(u^{It}(r))$ and $\exp(u^{In}(r))$ can be factored out and the sum of nodal diagrams $N^{I}(r)$ is the same in both quantities. This is strictly valid only when spatial overlap of those correlation factors vanishes, which happens when the fluid clusterizes.

Within these asumptions we can calculate radial distribution function of the homogeneous Bose gas by minimizing the total energy without the impurity (see Ref. [\cite{bosegas}]). The impurity distribution functions, $g^{It}(r)$ and $g^{In}(r)$, can then be solved by minimizing the chemical potential of one charged impurity, \cite{SKJLTP,MixMonster,EHLcluster}
\bea
\label{chemiIM2}
      \mu^{IM}&=& n_0\int d^2r\Biggl[ -\frac{e^2}{\varepsilon} 
      \frac 1 r\left(g^{It}(r) + g^{In}(r)-1 \right) 
 +\frac{\hbar^2}{2 m}|(\nabla\sqrt{ g^{It}(r)}|^2
 +|\nabla\sqrt{ g^{In}(r)}|^2)\Biggr]
\cr
 &+&\frac{1}{2}\int \frac{d^2k}{(2\pi)^2 n_0} 
	S^{IM}(k)~\tilde w^{IM}_{\rm ind}(k) -M E_{\rm bin}\,,
\eea
with the constraints of Eqs. (\ref{norm}).
In the last line we introduce the induced potential which collects the many-body correlation effects between the impurity and bosons.
\be
	\tilde w^{IM}_{\rm ind}(k) =
        \frac{\hbar^2 k^2}{4m} S^{IM}(k)\left[\frac 1{(S^{bb}(k))^2}-1\right]\,.
\label{inducedIM}
\ee
The set of  Euler equations which minimize $\mu^{IM}$ can be written as,
\bea
	-\frac{\hbar^2}{2 m} \nabla^2\sqrt{g^{It}(r)}+\left[-\frac{e^2}{\varepsilon r} 
	+w^{IM}_{\rm ind}(r)\right]\sqrt{g^{It}(r)}&=&E_{bin}\sqrt{g^{It}(r)}
	\cr
		-\frac{\hbar^2}{2 m} \nabla^2\sqrt{g^{In}(r)}+\left[-\frac{e^2}{\varepsilon r} 
	+w^{IM}_{\rm ind}(r)\right]\sqrt{g^{In}(r)}&=&0
\label{eulerIM}
\eea
with the Fourier transform $w^{IM}_{\rm ind}(r)$ of the induced interaction (\ref{inducedIM}). Equation (\ref{eulerIM}) defines the effective, coordinate space interaction 
\be
V_{\rm eff}(r) = -\frac{e^2}{\varepsilon r} +w^{IM}_{\rm ind}(r)
\ee
experienced by bosons around the bubble. 
If we think Eqs. (\ref{eulerIM}) as Schrodinger like equations then the fist equation gives the bound state solution in that effective potential and the second one is the zero energy solution with the boundary condition $\lim_{r\ra\infty}g^{In}(r)=1$.
However this set of equations is not that simple, because the induced potential contains nonlinear terms. A consistent solution is found by iteration and it converges within a few iteration cycles.
\begin{figure}[tb] 
\includegraphics[width=0.4\textwidth]{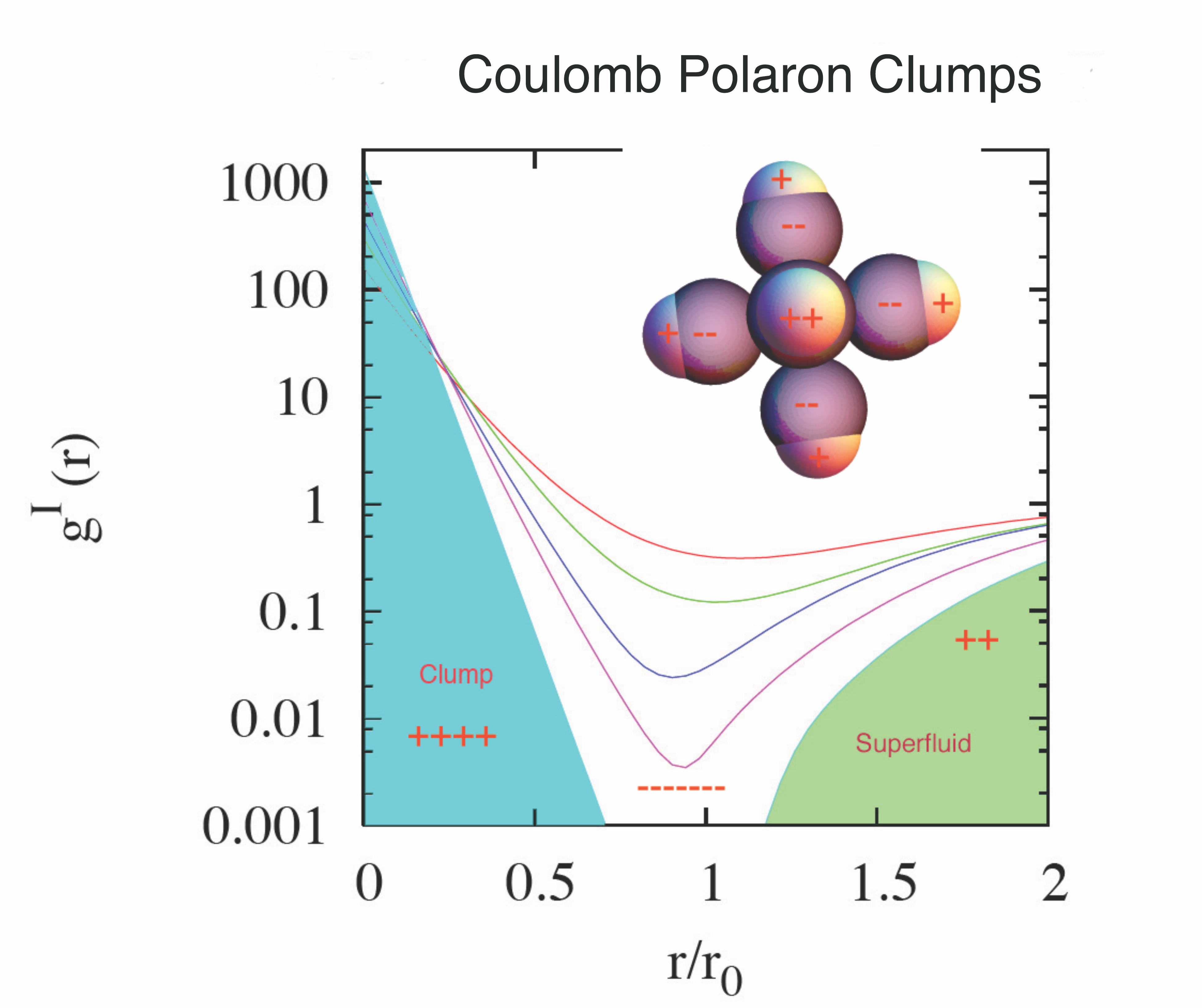}
\includegraphics[width=0.4\textwidth]{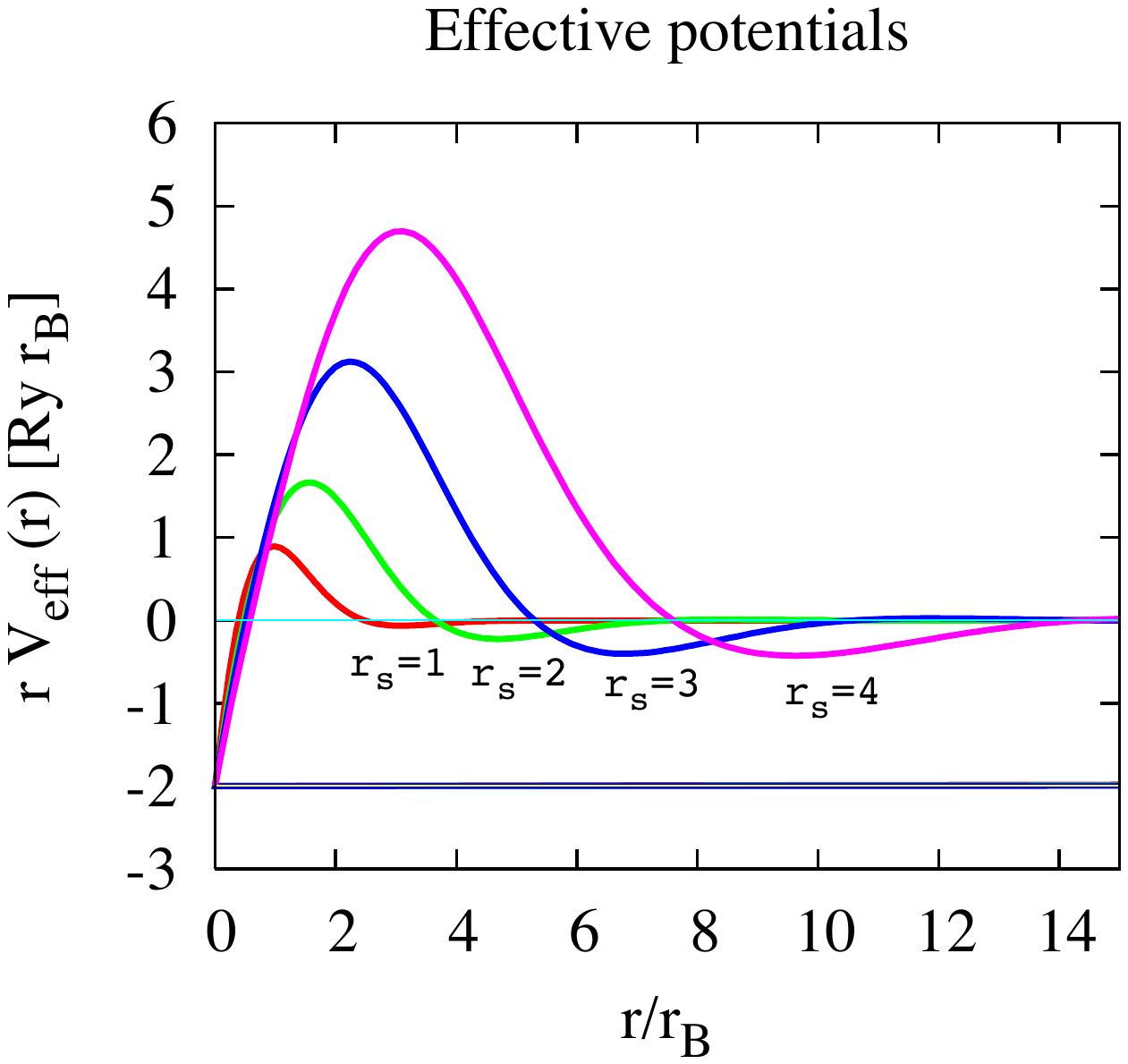}
\caption{(Color online) On the left is shown the distribution of dense bose fluid around an impurity $g^I(r)$ in logarithmic scale at different $r_s$ values for the range $r_s$= 2.5, 3, 3.5, 4 and 5. The minimum decreases with with increasing $r_s$. The density distribution is normalized to unity far away from the impurity. At the  value $r_s=4$ six particles are trapped into the clump and at the value $r_s=5$ seven particles are trapped. The value $r_0$ indicates here an average distance between particles. \newline
On the right are shown the effective potentials multiplied by $r$ for $r_s$ values indicated in the figure The constant line at -2 is the bare Coulomb potential in atomic units. The behavior of the effective potential shows a huge over-screening which increases with $r_s$ Of course the form of clusters should match the geometry of the Cu-O plane, then each cluster may be decomposed on elementary bricks as
that presented on the insert, left figure.
} 
\label{Fig2} 
\end{figure}
By studying the boson-impurity pair distribution function, $g^{In}(r)$, which describes the spatial charge-distribution of bosons around a single impurity polaron, we demonstrate the formation of CBs when the density decreases (see Fig. \ref{Fig2}.).  The oscillations in $g^{In}(r)$ are very similar to those presented schematically in Fig. \ref{Fig1}. One can clearly resolve the giant oscillation, the GCDO, closest to the impurity polaron where the high density area is surrounded by lower charged density. The amplitude of this oscillation increases with $r_s$. At the value $r_s=3.5$ this amplitude is so large that the charged fluid separates into two orthogonal phases. A small number of bosons or holes are trapped around the impurity and decoupled from the rest of the superfluid forming CB, because the strong repulsive over-screening of the Coulomb interaction prevents tunneling from the trapped states to continuum states. The effective potential multiplied with $r$ is also shown in Fig. 2. In the used atomic unit the pure Coulomb potential is constant -2 and as shown the indused many-body effects strongly over-screen at intermediate ranges the bare potential.  On insert, above we present an elementary bricks from which larger clusters are formed.

The results plotted in Fig. 2 are for the case where the impurity is in the two-dimensional plane and its charge is equal but opposite to the boson charge. They are very typical and similar ones are obtained for cases where the impurity is outside the superconducting layer or its charge is a multiple or a fraction of the boson charge.  The critical $r_s$ value depends on these conditions. For instance if the impurity charge is $-1e$, i.e. minus half of the boson charge, then the critical $r_s \sim 6.2$, or if the distance from an impurity to the superconducting layer is one Bohr radius then the critical $r_s \sim 12$. 

The dependence of the chemical potential of Coulomb bubbles having  different number of particles trapped on $r_s$ is presented in Fig. 3. The more particles are trapped into the bubble the lower chemical potential it has. From the Euler equations (\ref{eulerIM}) with get also the binding energy per trapped particle into the Coulomb bubble. It increases  with $r_s$ shown in Fig. 3. and has the limiting value -4 Rydbergs of the two dimensional Hydrogen like atom. The mutual repulsion reduced the binding energy when more particles are bound.

\begin{figure}[tb] 
\includegraphics[width=0.5\textwidth]{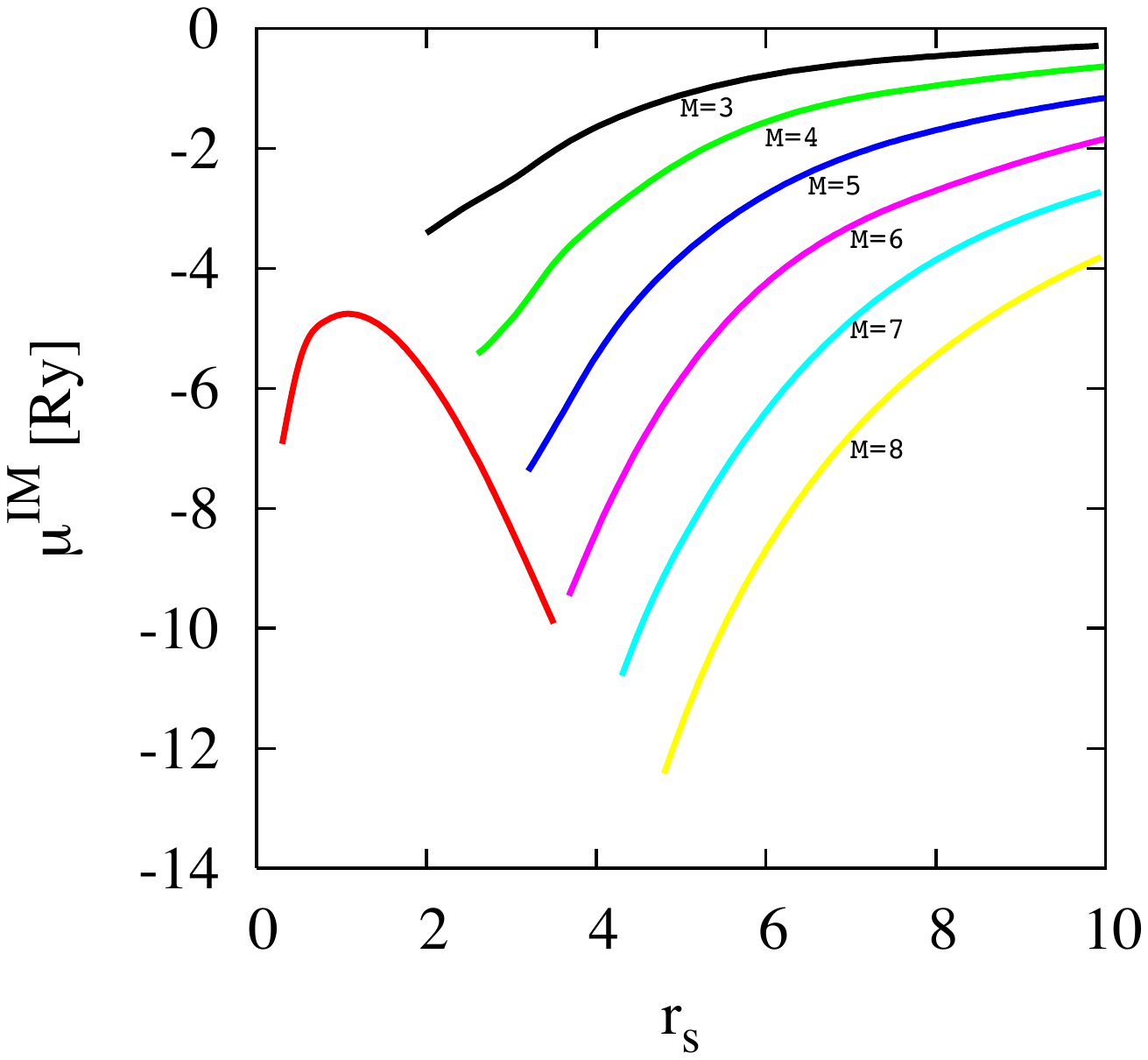}
\includegraphics[width=0.3\textwidth]{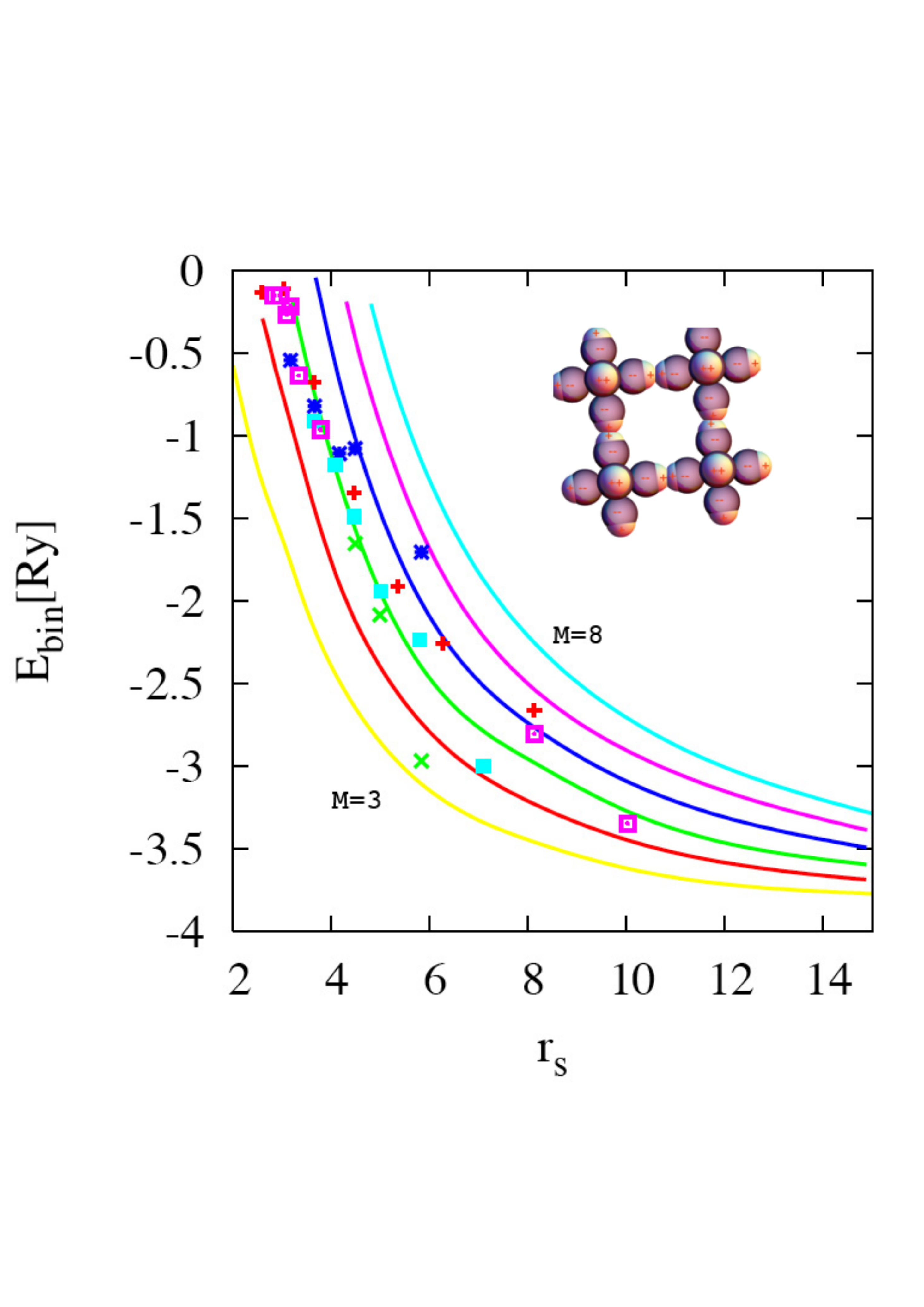}
\caption{(Color
online) On the left  we show chemical potentials as a function of $r_s$ for  Coulomb bubles or clumps having different number of particles trapped as indicated in the figure.  The more particles are trapped into a clump the lower the chemical potential becomes.\newline
In the right figure the binding energy/particle into the Coulomb bubble is shown as a function of $r_s$ for different number of particles $M$=3,4,5,6,7 and 8. The bottom yellow curve corresponds to $M=3$ particles trapped, while the top (cyan) curve  corresponds to $M=8$ trapped particles.  
We plot also here the experimental data  used by Gorkov and Teitelbaum (GT)\protect\cite{34,GT}) to fit  the resistivity and the Hall coefficient in LSCO\protect\cite{[21],[25]},  ARPES \protect\cite{[27]} and STM  experiments\protect\cite{[2]}. The ARPES data  are presented by red crosses.  The  GT fitting data taken from the $ab$ resistivity and the Hall
coefficient in LSCO\protect\cite{[21],[25]},  are presented by squares (cyan and magenta colors).  The rest crossed points correspond to STM data\protect\cite{Davies-2007,Davies-2008,[2]}.
On insert, right figure, we present a possible shape of the 8 particle cluster, which can comprise up to 4 CuO plaquettes. 
 \label{Fig3} } 
 \end{figure} 

\section{Comparison with Experimental ARPES and Hall Effect Data}

Gorkov and Teitelbaum (GT)\cite{34,GT}  have made a
detailed analysis of experimental data, which include   the temperature dependence both for resistivity and the Hall
coefficient in LSCO\cite{[21],[25]}, the ARPES experiments\cite{[27]} and STM\cite{[2]}. 
The Hall data were available for the whole range of concentrations and temperatures up to 900 K \cite {[21],[25]}.  By the analyses of these data they have shown  that there are  two components of charge carriers in cuprates.
In particular, Gorkov and Teitelbaum (GT)\cite{34,GT}  have found that the density of current carriers
in cuprates in a broad range of doping $x$  and temperatures $T$ is satisfied to the expression
\be
n(x) =n_0(x)+ n_1 \exp (-\Delta(x)/T)
\ee
where $n_0(x)=x$ for small $x$, the value $n_1$ is a temperature and doping independent and is equal approximately to $n_1 \sim 2.8$. It was also found there that $\Delta(x)$ is an universal function
of doping $x$. This  universal function $\Delta(x)$ fits well numerous experimental data in LSCO, both taken from Hall effect studies\cite{[25]} and from ARPES \cite{[27]} . GT have also argued that in order to fit numerous experimental data of STM studies
presented by Davis group\cite{Davies-2007,Davies-2008,[2]} the value of $\Delta(x)$ must be associated with the activation energy of some localized states, which may introduce inhomogeneity into these systems.
The GT  function $\Delta(x)$ in the equation for  the current carriers concentration may correspond to the binding energy of $M$ particles self-trapped into the Coulomb bubble, that is 
$\Delta(x) = -E_{bin}(x) $. In this case for the Coulomb bubbles with different number of particles trapped the concentration of the current carriers must satisfy to the relation
\be
n(x) =n_0(x)+ n_1 \exp (E_{bin}(x, M)/T)
\ee
where $M$ is the number of particles self-trapped into the Coulomb bubble.
Exactly in the same way as Gorkov and Teitelbaum  we have taken instead of the GT's  function $\Delta(x)$ presented in Refs.\cite{34,GT},  the binding energy of $M$ particles self-trapped into the Coulomb bubble, $-E_{bin}(x) $ and  plotted this function as a function of $r_s$, see Fig.\ref{Fig3}. for different number of particles $M$ trapped into the bubble. The lowest yellow curve corresponds to the bubble with 3 particles, while the top curve corresponds to the Coulomb bubble with $M=8$ particles trapped.
For intermediate curves (red, green, blue and purple) the number of particles increases from $M=4$ to $M=7$, respectively. 

For a comparison we plot on the same Figure the experimental data which were used by Gorkov and Teitelbaum to fit  the resistivity and the Hall
coefficient in LSCO\cite{[21],[25]},  ARPES \cite{[27]} and STM  experiments\cite{[2]}. There by red crosses the ARPES data  are presented. These data  perfectly coincide with theoretical curves (see, green and blue curves on the Figure) associated with the Coulomb bubbles having $M=5$ and $M=6$ particles. For small values of $r_s$ the bubbles are having $M=5$ particles, while with increasing $r_s$ further, ie for $r_s>5$  the bubbles will have $M=6$ particles. This is exactly following the prediction of our theory.  The  GT fitting data taken from the $ab$ resistivity and from the measurements of the Hall
coefficient in LSCO\cite{[21],[25]},  and presented by squares on the Fig. \ref{Fig3} (cyan and magenta colors) are consistent with the Coulomb bubbles having 5-6 particles self-trapped. In fact at relatively small $3<r_s <5$ these data correspond to  bubbles  having $M=5$ particles, while at larger $r_s$  there are more fluctuations in these data in a comparison with the calculated curves. The rest crossed points correspond to the STM data\cite{[2]}, which are also perfectly consistent with the Coulomb bubbles picture and numerical calculations, see the Fig. \ref{Fig3}. Thus the binding energy of the Coulomb bubbles agrees perfectly with the experimental data for the pseudogap in LSCO, extracted by GT \cite{34,GT}. Indeed, we have shown here that the GT function $\Delta(x(r_s))$  perfectly coincides with the function $-E_{bin}(r_s)$, which is the binding energy of  $M$ particles forming the Coulomb bubble in Eqs. (\ref{eulerIM}). This comparison with the  experimental data extracted by  GT\cite{34,GT} indicates that with increasing $r_s$ the average number of particles in the Coulomb bubble  increases from $M=5$ to $M=6$. It is important to note that in underdoped cuprates  Coulomb bubbles will always arise with different number of particles trapped. This is consistent with STM data\cite{Davies-2007,Davies-2008,[2]}, where various nanoscale superstructures in HTSC have been observed as well as with
transport data\cite{Anna-2008,Hussey-2009}.

\begin{figure}[tb] 
\includegraphics[width=0.3\textwidth]{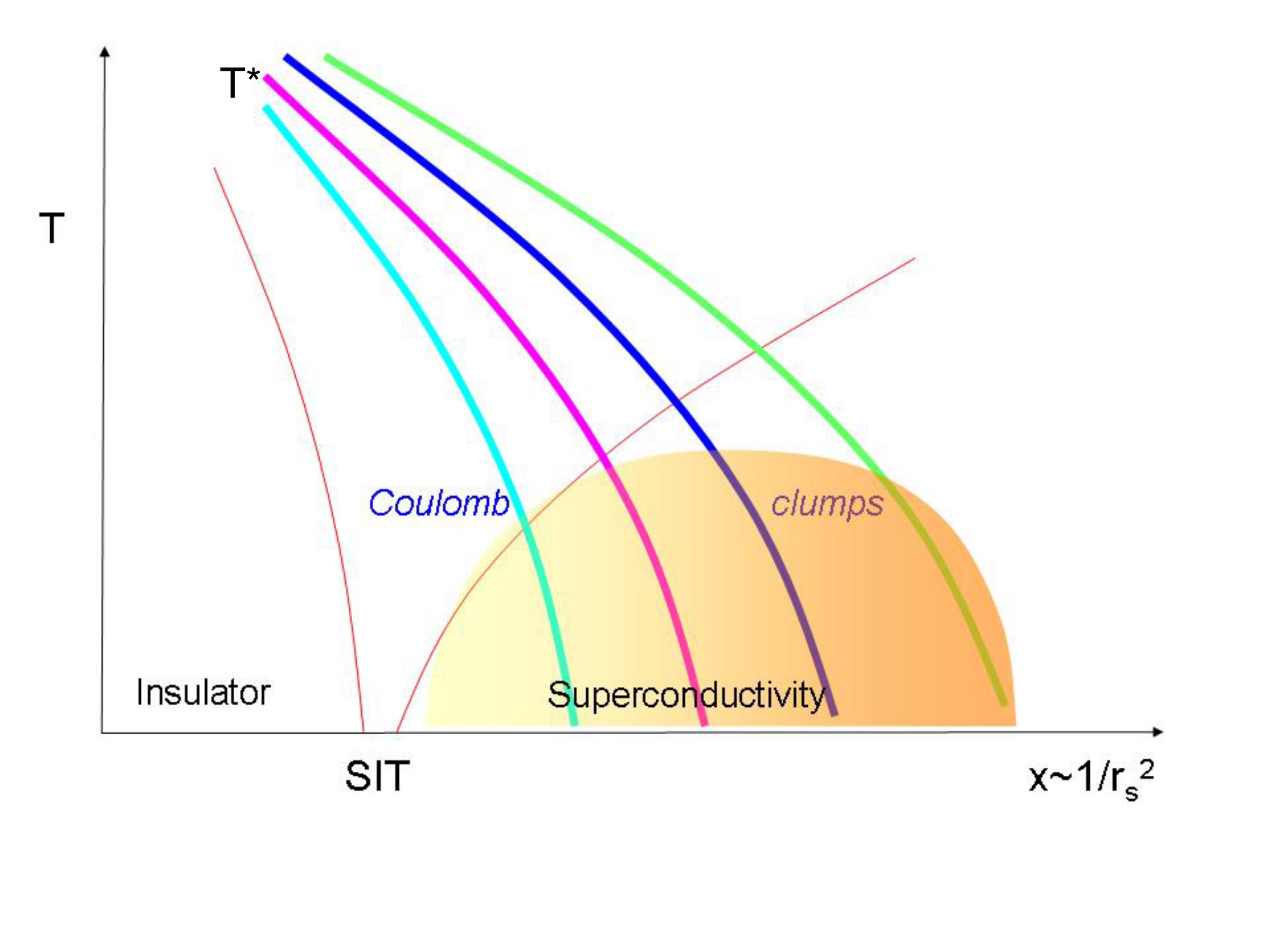}
\includegraphics[width=0.3\textwidth]{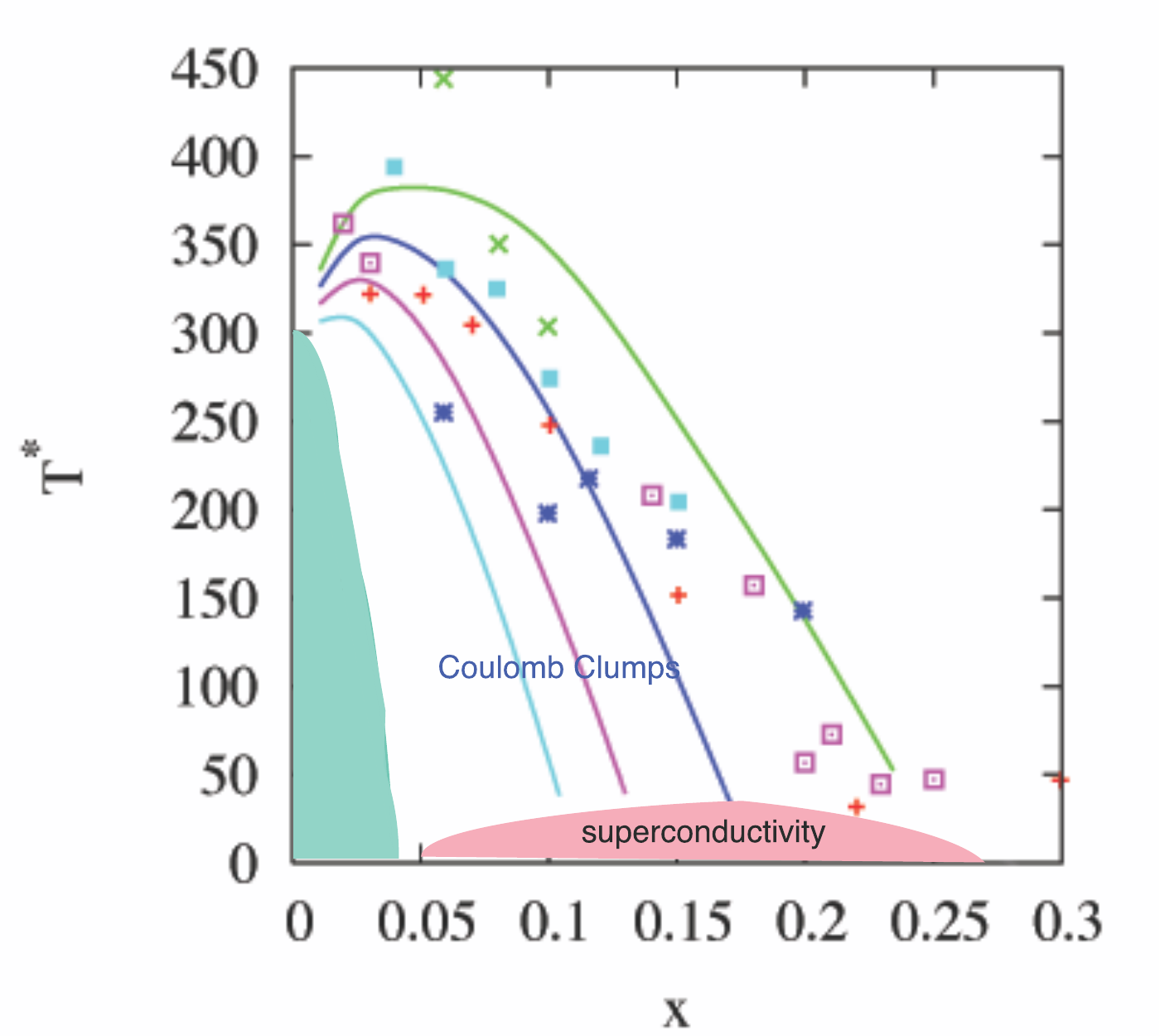}
\caption{(Color
online) On the left a schematic phase diagram, which emerges from the first principle microscopic many-body studies of the charged fermionic and bosonic liquid with embedded charged impurities or heavy small polarons.  The superconducting dome is noted by orange color, the stronger  orange color corresponds to larger concentration of the superconducting current carriers. The bold rain-bow color  lines corresponds to the pseudogap critical temperature. The pseudogap is determined as the binding energy of particles forming the Coulomb clump. The different colors of the rain-bow corresponds to the different number of particles trapped into the Coulomb clumps, ie the green curve corresponds to the dissociation temperature  for  the  clump consisting of 4 particles trapped.
The low cyan  curve corresponds to  a dissociation temperature for a clump with 7 particles.  The whole superconducting dome  represent a coexistence region of Coulomb clumps  with superconductivity. Inside the coexistence region there are superconducting microscopic clumps of different size randomly distributed.  This agrees with an analysis of the transport experimental data
\protect\cite{Anna-2008,Anna-2009,Hussey-2009}. On the right Figure we present a comparison of the experimental data for a pseudogap temperature
 with the critical temperature of dissociation of the Coulomb clumps calculated with the variational many-body formalism. The resistivity and the Hall coefficient in LSCO data taken from Refs\protect\cite{[21],[25]},  ARPES - from \protect\cite{[27]} and STM  experiments - from \protect\cite{[2]}. The ARPES data  are presented by red crosses. The transport data are presented by squares (cyan and magenta colors).  The rest crossed points correspond to STM data\protect\cite{Davies-2007,Davies-2008}. This comparison indicates that  these $r-$space states of the Coulomb bubbles may be indeed  associated with the copper
oxide pseudogap states identified in numerous experiments.}
\end{figure} 

 The saturation  in $E_{bin}(r_s\rightarrow \infty)$, or  in $\Delta (x \rightarrow 0)$ is about $0.25 eV$, which, according to our microscopic calculations, corresponds to  $- 4 Ry =-2  m e^4/(\hbar^2 \epsilon_0^2)$. From this expression for the saturation of the binding energy at small doping or large $r_s$ we  can extract the static dielectric constant and compare this value obtained from other experiments.  For example if $- 4 Ry =-2  m e^4/\epsilon_0^2/\hbar^2 =0.25 eV$, and the hole charge  $e=1$ we can estimate that
$\epsilon_0=20$, when $m=2 m_e$,  however when we use for  effective mass its optical value for LSCO $m=4 m_e$  taken from the reference\cite{Padilla} we obtain that  $\epsilon_0=29.5$.  These values for $\epsilon_0$ are in a perfect agreement with independent experiments in  BSCO and LSCO where this value of dielectric constant has been measured directly \cite{Chen}. 

This comparison with experimental data  indicates that the GT finding of the activation energy $\Delta(x)$ is consistent with our binding energy, $E_{bin}(x)$, for the whole doping range where both these functions exist. Thus our results reinforce the  conclusion deducted by Gorkov and Teitelbaum from an analysis of numerous experimental data\cite{34,GT} that there are two types of current carriers in cuprates: doped,  associated with the function $n_0(x)\sim x$ and activated associated with activation energy $\Delta(x)$, which is nothing but an activation energy of electron or holes bound in the Coulomb bubbles. GT have also found that the pseudogap critical temperature $T^*$ fits well to the equation
\be
T^*(x)=-\Delta(x)/\ln(x)
\ee
Following our finding and identification of $E_{bin}(x) =\Delta(x)$ we can perform a microscopic estimation of the $T^*$ temperature, ie
\be
T^*(x)=-E_{bin}(x)/\ln(x)
\ee 
Our schematic phase diagram in the plane defined by the temperature and concentration of dopants is shown in Fig. 4, left figure. We also present there on the right Figure, the dependence of  $T^*(x)$ both our theoretical estimations (the solid colored curves) and various experimental data collected by Gorkov and Teitelbaum\cite{34,GT}.
Following again arguments given by GT, such values of $T^*$ calculated in our model are in a perfect agreement with existing experimental data on $T^*$ \cite{34,GT}. It is very probable that these  Coulomb bubbles  are associated with vanishing of the van Hove points on the Fermi surface observed in various ARPES\cite{[27]}. These data need more analyses in the framework of the Coulomb bubbles picture and will be published elsewhere\cite{Kus-Saarela-2009}.
 
Below each rainbow color lines(see, Fig. 4.) the bubbles exist, and this is valid both inside the superconducting dome indicated by the orange color and in the insulating and metallic phases. The coexistence region of the superconducting fluctuating phase consisting of  insulating and metallic islands is located between red thin curves on the Figure. Below the superconducting transition (inside the orange dome, see, Fig. 4.) the density of CBs is nearly the same as above the transition. Of course due to activation process the number of bubbles with smaller number of particles increases with temperature until all bubbles will evaporate.
 Below the transition we have bubbles and the superconducting condensate while above the transition there are the CBs and free bosons (see, Fig. 4.). 
We assume here that the condensate fraction of bosons vanishes at $T_c$. With increasing temperature more and more free bosons evaporate from CBs. We associate the temperature, at which all bosons have evaporated from the CBs (the energy of their complete ionization), with the critical temperature, at which the pseudogap state disappears.  The rainbow colour solid lines  in Fig. 4. represent the schematic(left) and calculated(right figure)  dependence of the pseudogap critical temperature $T^*$ on doping (see, also Ref.\cite{[20]} for details). The different lines of the rainbow are associated with the critical temperature of the bubbles dissociation having a different fixed number of particles trapped. 
 
To summarize this section on a comparison with the GT fitting data\cite{34,GT} we have obtained that with decreasing doping from the overdoped state at the optimal doping  arises a splitting of the broad band into two subbands. One of them remains always a broad one and the density of the current carriers  there weakly depends on the temperature and doping $n_0(x)=x$. The second band is very narrow or just flat and arises due to a Coulomb instability  and formation of the Coulomb bubbles and electron-hole strings\cite{Bianconi-Kusmartsev}. The gap separating these bands is associated with the binding energy of electron and holes forming the Coulomb bubbles, which also induce nanoscale inhomogeneities in these systems. Thus our results consistent with GT ones \cite{34,GT} state  that there are two types of sources for current carries: 1) from localized "bubbles"
and 2) just free carriers  arising due to a direct doping effect. We have also found that similar situation arises in YBCO samples, where analogous gap, $\Delta(x)$, has been identified in optical experiments\cite{Yu,Mihailovic}. There in heavily underdoped samples the resistivity and optical conductivity has similar activation character associated with the quantity $\Delta(x)$  which is in our picture associated with $ E_{bin}(x,M)$. It is even more striking that this function
is very similar for all cuprates, not only qualitatively but also quantitatively defined as the binding energy of the Coulomb bubbles.

\section{Coulomb polaron clumps embedded in HTSC}\label{HTSC}

In numerous Scanning Tunneling Microscope (STM) experiments  [3-6] various nanoscale structures have been observed in the superconducting state both in underdoped and in optimally doped cuprates.  The appearance of these structures depends on the value of the bias voltage used in STM experiments. At small values no such structures are visible. However, when the bias voltage is larger than some critical value these structures are unveiled. This fact has been intensively discussed in the literature and considered as the main puzzle of HTSC, which can not be explained by any of existing theories \cite{11}. 

This puzzling fact is very naturally described in the framework of our results. Each CB may be viewed as a quantum dot Ð a mountain located on the bottom of the deep sea of the superfluid charged liquid. When a bias voltage is applied the depth of the superfluid sea (a level of chemical potential) or the thickness of the superfluid density decreases. The effect is dramatic for areas where the CBs are densely packed. Then, at some critical bias voltage the superfluid density will be locally broken and the CBs begin to take part in the STM current and therewith become visible as nanoscale structures.

McElroy et al.\cite{11,Davies-2008} have identified populations of atomic-scale impurity states whose spatial distribution follows closely that of the oxygen dopant atoms. They also found a close connection between the nanoscale structure of the superconducting order parameter (electronic disorder) and locations of dopant atoms. This is obviously possible only if dopant atoms pin CBs. As stated before, CBs are like quantum dots in the Coulomb blockade regime. Then due to the large charging energy the low-energy spectral weight shifts to higher energies, and the superconducting coherence peak becomes strongly suppressed near impurities. At the same time the low-energy, condensate bosons scatter very weakly from the CBs. 
In Fig. 4 we present the comparison of our results for $T^*$ with available experimental data [\cite{11,34,GT,Davies-2008}].

 The Coulomb clumps discussed here are induced by oxygen impurities and small polarons and form a narrow or even flat band. This narrow  or flat band was originated due to a geometry of the Cu-O plane from one side where oxygen electrons or holes are bridging Cu atoms. Between Oxygen and Copper atoms there is a staggered field
 associated with the value $\epsilon_p>0$. Such staggered field together with the Coulomb interaction (namely, due to an over-screening of the Coulomb interaction) give rise to localized electron-hole strings or polaron clumps.  The formation of such strings
 are also supported by relatively strong electron-phonon interaction and that leads further to the narrowness of the electronic band. When doping decreases such string and clumps cover more and more of the Cu-O plane until the whole plane will be covered.
   These polaron strings and Coulomb clumps form a background on which the superconductivity is formed. The latter is of course associated with the broad band formed due to holes moved over Cu sites. Probably the mechanism of the superconductivity is related to an interplay between Coulomb clumps and mobile liquid of holes.
    Such speculation is consistent many experiments, in particular with ARPES \cite{[27]} and anomalous criticality in electrical resistivity data\cite{Hussey-2009}. 
    In first type of experments each emitted photoelectron leaves behind a hole surrounded by local lattice distortions associated with the electronic strings on oxygens. The energy of these lattice distortions  is about the deformation energy per electron per string. In this case  ARPES experiments show that lower energy is needed to break the bound state of  electrons and holes forming the Coulomb bubble than for ionisation of the electrons or holes in the thermal activation process. Indeed,  as it was also noticed by GT\cite{34,GT} this feature is clearly illustrated on the Figs. 3 and 4.
There  the ARPES data \cite{[27]} correspond to the lower binding energy, or lower pseudogap value.
The difference is about 20 $meV$ which is about the typical energy of the strings formation per electron, see for details Refs. [\cite{Kusmartsev-1999,Kusmartsev-2000,Kusmartsev-IJMP,Bianconi-1997,Bianconi-Kusmartsev}]. The pedagogical review about the string formation is presented in Ref. [\cite{Kusmartsev-2004}]. The importance of the
lattice distortions in the formation of the pseudogap has been noted by GT who have also speculated that the lattice effects are also involved in formation of some e-h structures, which we present here as e-h strings or Coulomb clumps. Indeed, in the present paper we have found that e-h structures speculated by GT are, in fact, corresponding to the Coulomb bubbles or clumps which are naturally and originally formed due to over-screening of the Coulomb interaction when the density of the current carriers changes.  Additional factors such as the existence of the staggered field on the Cu-O plane and strong electron-phonon interaction reinforce our conclusion. 
GT have also noted that the binding energy does not go to zero (see, recent ARPES data\cite{34,GT,[27]}). It indicates that when Coulomb bubbles vanish (slightly above above the doping $x\sim 0.27$) the strings still remain since the energy gap still persists and that is about the energy of local lattice distortions existing in the oxygen electronic strings, ie it is about 20 $meV$.  In the second group of experiments\cite{Hussey-2009} they found that when the carrier number falls the
effective interaction responsible for the (anisotropic)
T-linear scattering term in electrical resistivity becomes progressively
stronger. We found that when the carrier number falls the
over-screening of Coulomb interaction increases that leads to Coulomb clump formation. 
Thus, this is a consistent with our conclusion about the Coulomb clump formation.
Moreover, they found that this term closely correlates with  superconducting
critical temperature $T_c$ and the condensation energy. 
The Coulomb clumps  described here  may play a role of scatterers responsible for a linear T- resistivity analyzed in the Ref. \cite{Hussey-2009}.
The important conclusion made in this paper is  that the region of the linear T- resistivity increases when temperature decreases. Moreover they claim that such scatteres, which we ascribe to the Coulomb clumps do exist even in strongly overdoped regime where there is no superconductivity. So in the framework of our theory
 it is possible that bubbles arise before (ie at higher doping) the superconducting state emerges.

\section{Conclusions}

Recently, the most of the attention both of experimental and
theoretical communities has been addressed to the phase separation in the
underdoped regime between a first undoped antiferromagnetic  phase
and a second doped metallic phase of cuprates.  However, in majority of these papers
the role of the long-range Coulomb interaction has not been properly addressed. On a contrast with all these papers we have shown that the Coulomb force is the major force driving the microscopic electronic phase separation.
Moreover,  we have presented here an emerging theoretical scenario of microscopic phase separation driven by the long-range Coulomb forces.
This scenario grabs key physical aspects of the phase diagram of cuprates, see, Figs 3 and  4. It was shown that the  model of strongly correlated charged Bose fluid is appropriate for the superconducting phase of all cuprates.  The highest critical temperature of the superconducting transition in cuprates
is attained within the phase-separated state, where the first microscopic quantum droplets of electronic inhomogeneities (Coulomb bubbles coexisting with electronic strings) arise.

The recent discovery of high temperature superconductivity in FeAs multi-layered
materials \cite{F1,F2,F3,F4,F5,F6,F7,F8,F9,F10,F11} provides another class of systems where the
Coulomb interaction and the two component physics of the electron-hole quantum liquids discussed above do exist. The matter of fact some of these undoped
FeAs  compounds may correspond the state known as excitonic insulator.  However  the difference with the case of cuprates discussed above is the ratio of the electron and hole masses, which is not as large as in cuprates.  The fact that  in cuprates  electrons are localised means that their masses tend to infinity (due to the polaron effect). In contrast to that the fact that in these FeAs based materials the effective hole and electrons masses
may be compared leads to a large variety of possibilities which arise with the electron or hole doping of these excitonic insulators. Although  there exists a similarity of the two-component physics in cuprates
 and the formation of Coulomb bubbles (and therewith the formation of the pseudogap state) the details of such formation may be different. We believe that the classification of such pseudogap states of the excitonic insulator  and the nature of the superconductor-insulator transitions in these Fe-based compounds will be established in a near future.

\section*{Acknowledgments}
The work was supported by the ESF network-program AQDJJ and European
project CoMePhS.

\end{document}